\documentclass[runningheads]{llncs}
\usepackage{graphicx}
\usepackage{hyperref}
\usepackage{amsmath}
\begin{document}

\title{Graph Attention Network based Pruning for Reconstructing 3D Liver Vessel Morphology from Contrasted CT Images}
\author{Donghao Zhang \inst{1}, Siqi Liu \inst{2}, Shikha Chaganti \inst{2}, Eli Gibson \inst{2}, Zhoubing Xu \inst{2},\\ Sasa Grbic \inst{2}, Weidong Cai \inst{1}, Dorin Comaniciu \inst{2}}
\institute{\textsuperscript{1} School of Computer Science, The University of Sydney,\\ Sydney, NSW, Australia \\ \textsuperscript{2} Digital Technology \& Innovation, Siemens Healthineers,\\ Princeton, NJ, USA}

\titlerunning{Graph Attention Network Pruning}
\authorrunning{Zhang et al}
\maketitle

\begin{abstract}
With the injection of contrast material into blood vessels, multi-phase contrasted CT images can enhance the visibility of vessel networks in the human body. Reconstructing the 3D geometric morphology of liver vessels from the contrasted CT images can enable multiple liver preoperative surgical planning applications. Automatic reconstruction of liver vessel morphology remains a challenging problem due to the morphological complexity of liver vessels and the inconsistent vessel intensities among different multi-phase contrasted CT images. On the other side, high integrity is required for the 3D reconstruction to avoid decision making biases. In this paper, we propose a framework for liver vessel morphology reconstruction using both a fully convolutional neural network and a graph attention network. A fully convolutional neural network is first trained to produce the liver vessel centerline heatmap. An over-reconstructed liver vessel graph model is then traced based on the heatmap using an image processing based algorithm. We use a graph attention network to prune the false-positive branches by predicting the presence probability of each segmented branch in the initial reconstruction using the aggregated CNN features. We evaluated the proposed framework on an in-house dataset consisting of 418 multi-phase abdomen CT images with contrast. The proposed graph network pruning improves the overall reconstruction F1 score by $6.4\%$ over the baseline. It also outperformed the other state-of-the-art curvilinear structure reconstruction algorithms.
\end{abstract}

\section{Introduction}
With the injection of contrast material into blood vessels, multi-phase contrasted CT images can enhance the visibility of vessel trees in the human body.
Reconstructing the 3D geometric morphology of the liver vessels from such contrasted CT images can enable multiple computer-aided preoperative surgical planning applications, such as 3D visualization, navigation, and blood-flow simulation.
It is important to use computer-aided systems to automate the delineation of 3D liver vessels from CT images since it is a highly labor-intensive process in clinical practice.
However, it remains a challenging problem due to 
(1) the required integrity of the 3D vessel reconstruction is high for applications such as liver surgical planning since the reconstruction errors can bias decision-making process of the physicians;
(2) image noise and irrelevant anatomic structures sharing the similar intensity, making thresholding solutions impractical;
(3) the morphological variability of the liver anatomy making it hard to infer the vessel presence using prior knowledge;
(4) the difficulty of scaling up the voxel-wise labelled training dataset as in \cite{huang2018robust} due to the cost of annotating 3D vessel segmentation; and 
(5) depending on the imaging quality, the unknown number of liver vessel trees (portal vein, hepatic artery, and hepatic vein) expected to be visible in multi-phase contrasted CT.

Many previous curvilinear structure algorithms~\cite{rivulet2,app2,wu2011segmentation,zhao2018monocentric} rely on the accurate segmentation of the curvilinear structure.
The image processing based automatic curvilinear segmentation is achieved by designing hand-crafted curvilinear filters~\cite{chi2010segmentation,aylward2002initialization}.
By adapting the deep learning techniques, a few recent studies used convolutional neural networks (CNN) to replace the image filters for vessel segmentation~\cite{wu2018multiscale,wu2019vessel,zhang2018deep}. With the development of deep learning, many segmentation architectures are
proposed by improving the U-Net~\cite{ronneberger2015u}. The multi-scale refinement with the cascaded architecture~\cite{wu2018multiscale} improves vessel segmentation accuracy. The multi-path supervision and inception-residual blocks~\cite{wu2019vessel} are proposed to achieve a better performance. Beyond methods targeting at segmentation of 2D vessel images, 3D vessel segmentation methods~\cite{vessel3d,kitrungrotsakul2019vesselnet} were proposed.  Probability of centering voxel being vessel is predicted with three 2D orthogonal slices input and DenseNet as its backbone for classification network~\cite{vessel3d}.
However, such segmentation models are limited by the conventional convolutions in Euclidean space, neglecting the topological vessel connectivity.
To better model the vessel connectivity, the graph neural network (GNN) has been adapted into the image segmentation model~\cite{shin2019deep}.
There have been also a few early studies using GNN in medical imaging applications such as biomarker identification \cite{asd_gnn}, cerebral cortex parcellation \cite{parcellation} and disease-gene relation determination~\cite{han2019gcn}.
Many curvilinear structure reconstruction methods have also been proposed in the community of single-neuron reconstruction~\cite{rivulet2,app2,smarttracing,tremap,snake,neutube,most}.

In this paper, we propose a framework to reconstruct 3D vessel morphology from 3D multi-phase CT images by combining the fully convolutional neural network and the graph attention network.
We first train a vessel enhancement CNN to highlight the vessel centerlines. Based on the enhanced vessels, we use the recent 3D tree tracing algorithm \cite{rivulet2} to initialize the vessel graph tracing with high sensitivity and low specificity.
To prune the false-positive branches, we use a graph attention network with graph attention layers (GAT) to estimate the confidence of each sub-branch in the initial reconstruction.
We convert both the initial vessel reconstruction and the ground-truth reconstruction to their dual graph in which each node represents a sub-branch in the original graph.
The input graph nodes interpolate the CNN features from the 3D vessel enhancement CNN layers and use them as the input GAT features.
The output GNN network maintains the same graph topology while each node outputs the confidence that its corresponding sub-branch exists.
We evaluated the proposed framework on an in-house dataset with 418 3D abdomen multi-phase contrasted CT images.
Our results show that our method outperforms the baseline without graph attention network pruning by $6.4\%$ F1 score.
We also show that the entire proposed framework achieves the state-of-the-art compared to previous curvilinear structure reconstruction methods.

\section{Methods}

\begin{figure}[t]
\centering
\includegraphics[width=1\linewidth]{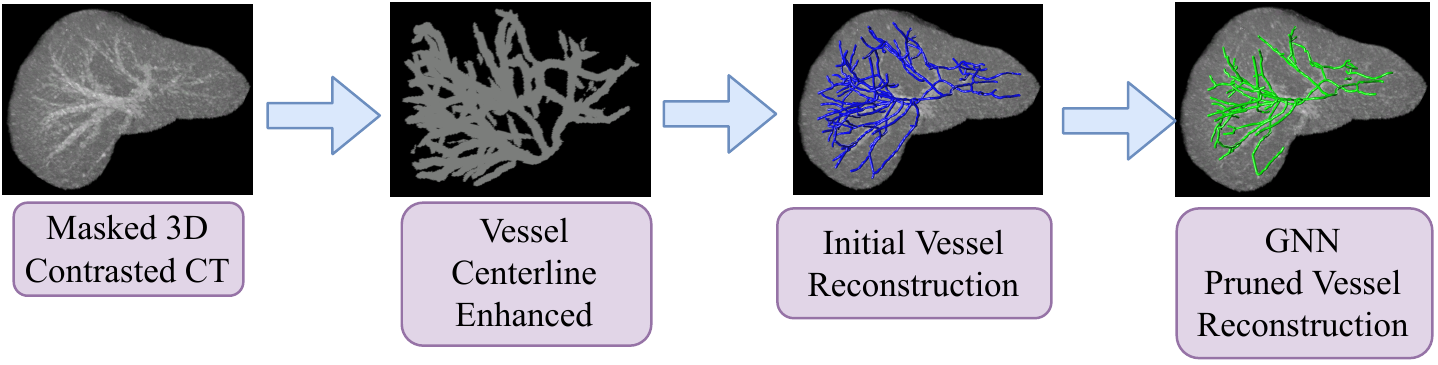}
\caption{The illustration of the overall liver vessel reconstruction framework.}
\label{fig:overall}
\end{figure}

\subsection{Liver Vessel Morphology Initialization}
As shown in Fig.~\ref{fig:overall}, the liver region is first cropped out using a trained liver segmentation model~\cite{di2in}.
We then train a 3D fully convolutional neural network to highlight the centerlines of the liver vessels.
Given vessel centerline models manually traced in 3D, the ground-truth centerline heatmap is generated as in~\cite{triple_cross}:
\begin{equation}
    d(p) =
    \begin{cases}
      e^{\alpha(1-\frac{D(p)}{d_M})}, & \text{if } D(p)\leq d_M\\
      0                             & \text{otherwise}
    \end{cases}
\end{equation}
where $D(p)$ is the perpendicular Euclidean distance from any 3D image coordinate $p$ to its closest ground-truth centerline;
$\alpha$ and $d_M$ are the decay rate and the heatmap radius, respectively.
$d(p)$ is normalized to $[0,1]$.
To train the CNN, we randomly sample CT vessel patches of size $64 \times 64 \times 64$ within the liver.
The online patch sampler balances the number of patches with vessels to be 5 times as the number of patches without vessels.
We use the binary cross entropy loss as the CNN training objective and optimize the parameters with AMSGrad-Adam~\cite{amsgrad}.
The training patches are augmented by flipping along 3 orthogonal directions.
We use the sliding window strategy with overlapping to predict the entire 3D CT volumes with online test-time augmentation by averaging the responses from 6 flips.
Only the central $32 \times 32 \times 32$ area of the sliding window output is written to the output volume. 
With the predicted centerline heatmap $d^\prime(p)$, we initialize the vessel morphology reconstruction using an algorithm originally used for tracing single neuron models from 3D light microscopy images \cite{rivulet2}.
We first dilate the binarized heatmap $d^\prime(p)$ to coarsely merge the disconnected components.
The tracing algorithm is applied on each connected component area of the continuous heatmap separately since there can be multiple vessel trees visible in the same multi-phase CT.

\begin{figure}[!tb]
\centering
\includegraphics[width=1\linewidth]{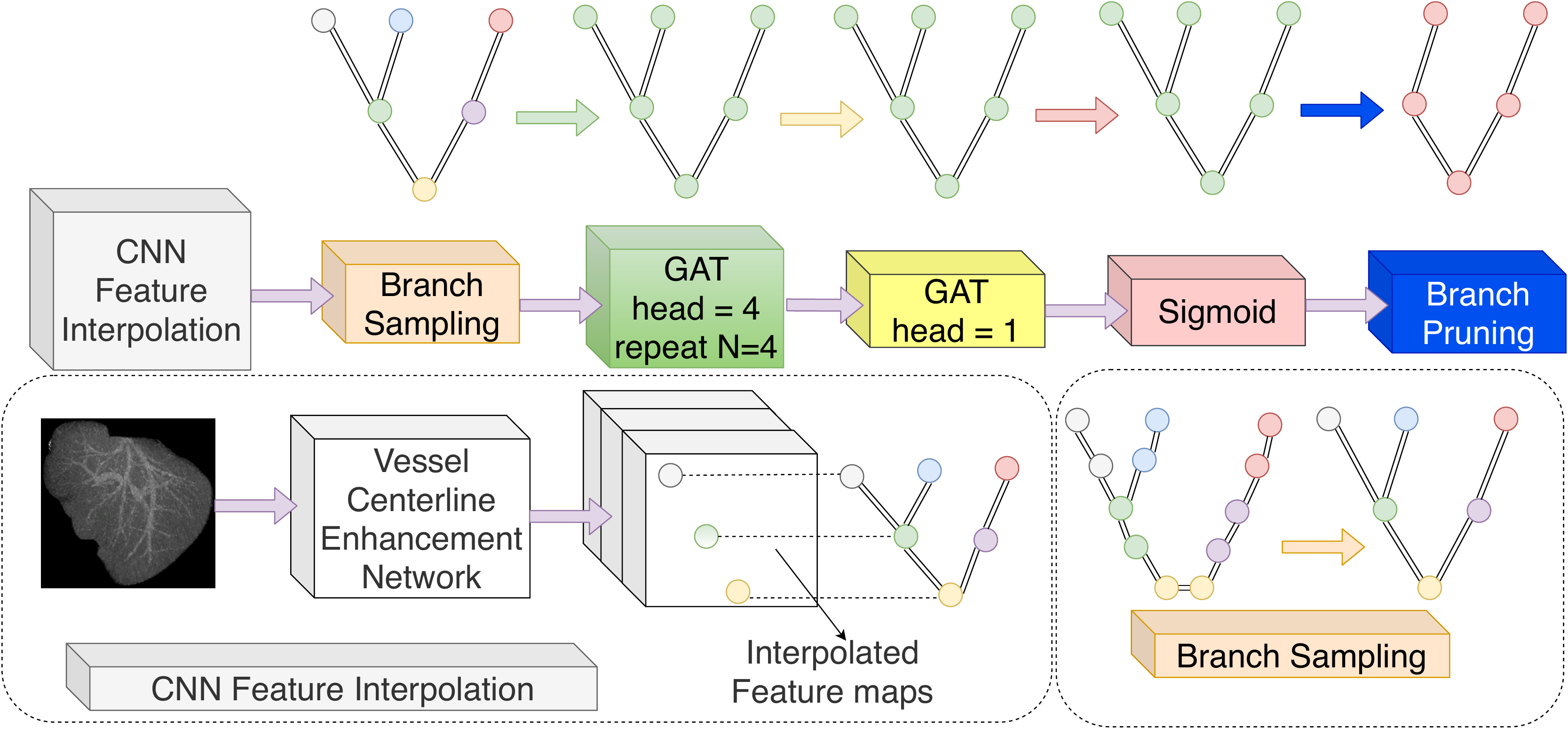}
\caption{The illustration of the proposed GNN-based vessel reconstruction pruning.}
\label{fig:gnn_prun}
\end{figure}
\begin{figure}[!htb]
\centering
    \begin{minipage}{0.36\linewidth}
      \centering
      \centerline{\includegraphics[width=1.\linewidth, height=0.6\linewidth]{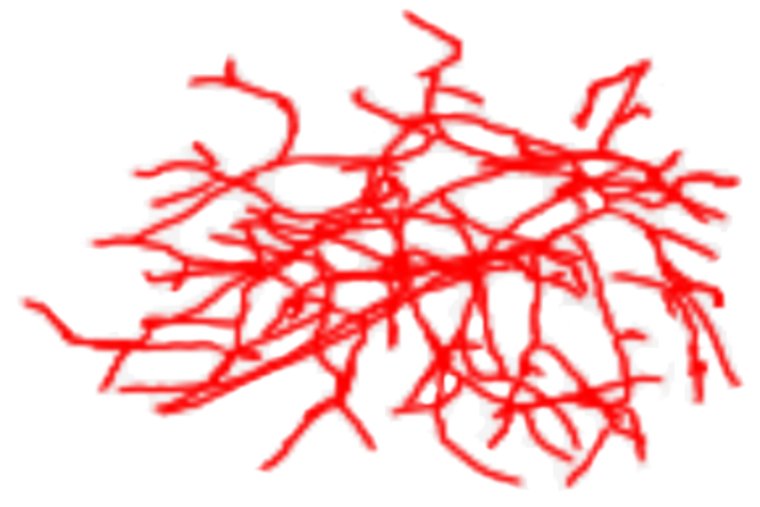}}
      \centerline{(a) Initial Tracing}
    \end{minipage}
    \begin{minipage}{0.36\linewidth}
      \centering
      \centerline{\includegraphics[width=1.\linewidth, height=0.6\linewidth]{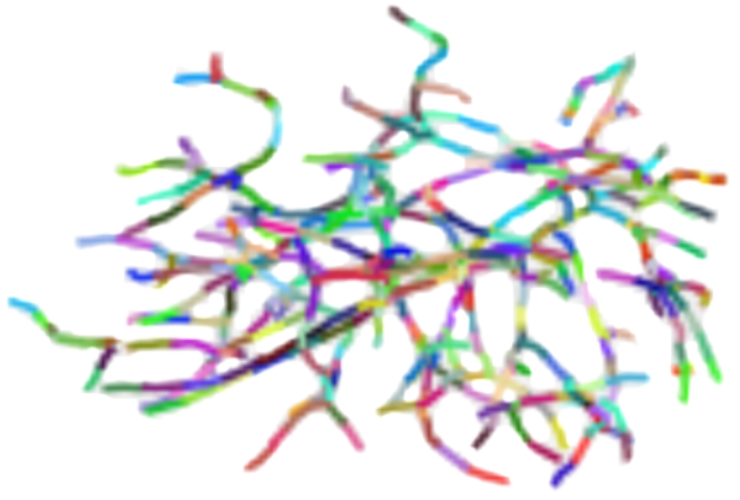}}
      \centerline{(b) Colored Branch Segments}
    \end{minipage}
    \centering
    \begin{minipage}{0.36\linewidth}
      \centering
      \centerline{\includegraphics[width=1.\linewidth, height=0.6\linewidth]{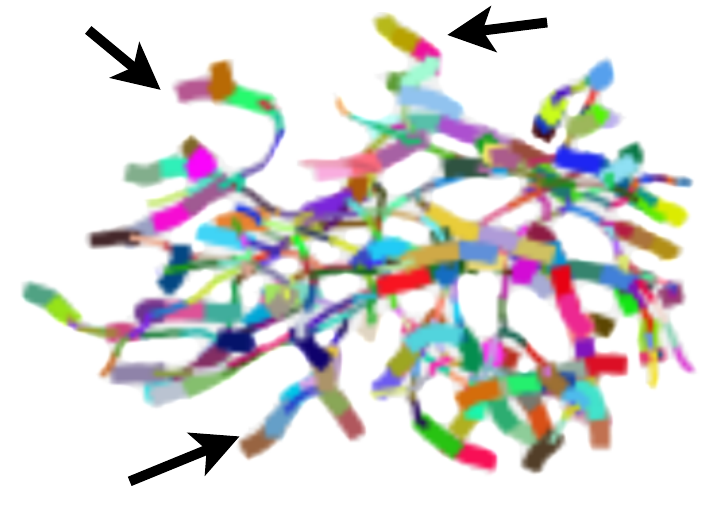}}
      {(c) Pruning Scores from GNN}
    \end{minipage}
    \begin{minipage}{0.36\linewidth}
      \centering
      \centerline{\includegraphics[width=1.\linewidth, height=0.6\linewidth]{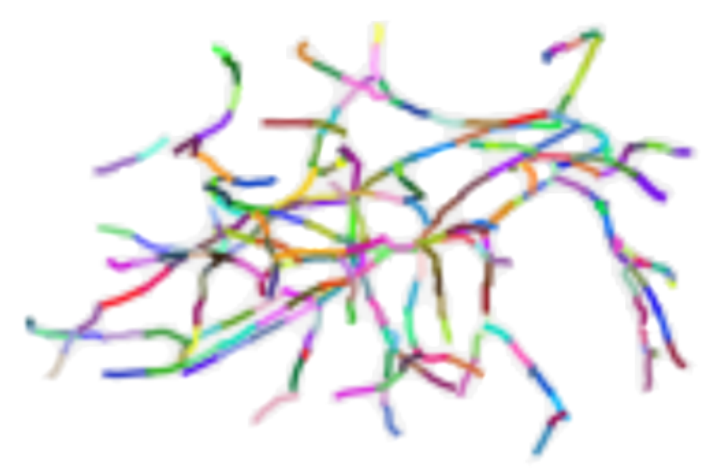}}
      \centerline{(d) GNN Pruned Tree}
    \end{minipage}
    \begin{minipage}{0.36\linewidth}
      \centering
      \centerline{\includegraphics[width=1.\linewidth, height=0.6\linewidth]{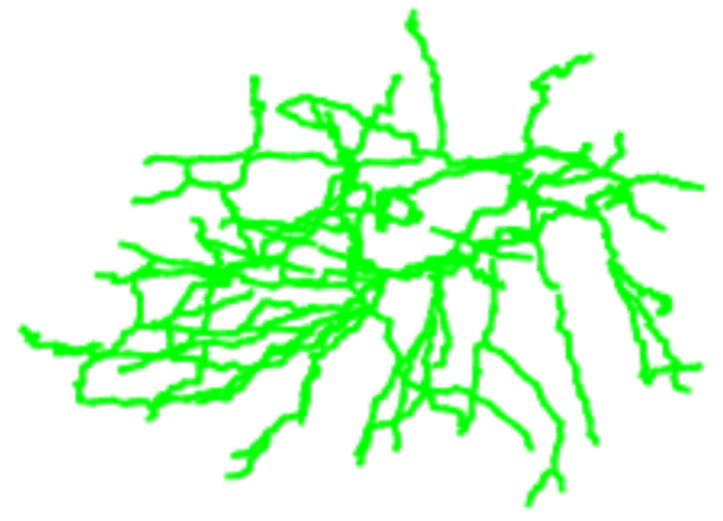}}
      \centerline{(e) Ground Truth}
    \end{minipage}
    \begin{minipage}{0.36\linewidth}
      \centering
      \centerline{\includegraphics[width=1.\linewidth, height=0.6\linewidth]{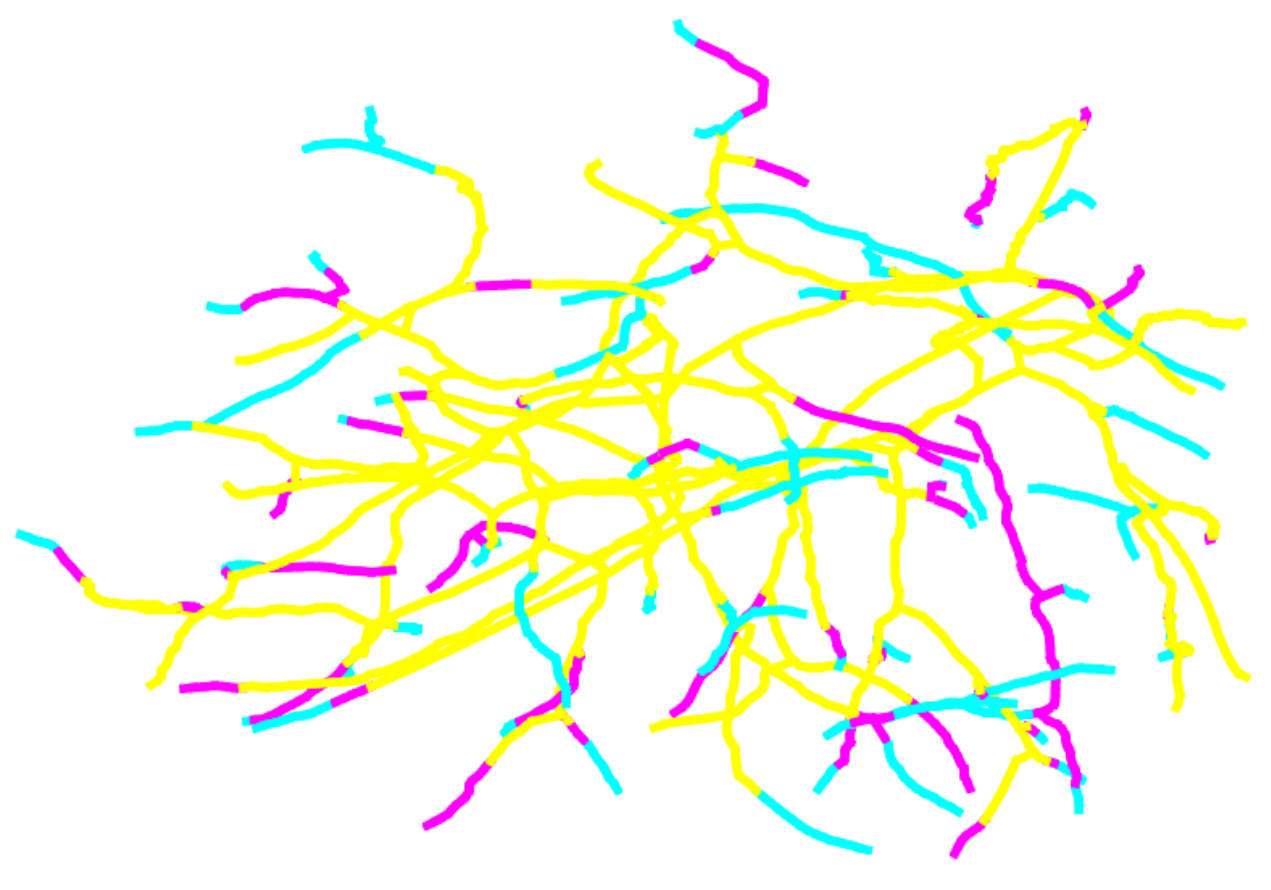}}
      \centerline{(f) GNN Pruned Branches}
      \label{fig:gnn-pruned}
    \end{minipage}
    \caption{Visualisation of the GNN pruning. (b) is the visualisation of the broken branch segments with different colors. (c) is visualization of GNN prediction and thicker branches have a higher probability of being pruned. (d) is the final vessel graph after discarding the predicted false positives. (f) compares the pruned graph with the ground-truth graph: Yellow represents true positives; Cyan represents false positives that are successfully discarded by GNN; purple represents the false negatives that are wrongly discarded.}
    \label{fig:gnn_prun_vis}
\end{figure}
\subsection{Graph Attention Network for Branch Pruning}
The initial tracing can produce many false-positive branches since the vessel enhancement network and the tracing algorithm are not jointly optimized.
Thus, we train a graph attention network to prune the false-positive branches by aggregating the CNN image features using the initial tracing graph as depicted in Fig.~\ref{fig:gnn_prun}.
We first break the initial tracing graph into branch segments $B_i$ with a length threshold $|B|$ as in Fig.~\ref{fig:gnn_prun_vis} (a) and (b).
The initial tracing graph is then transferred into its dual graph where each node represents its branch segment $B_i$ and the dual edges maintain the original topology.
For each node in the dual graph, we sample the CNN features from the first two and last two layers of the vessel enhancement CNN as $\Vert_{l=1}^{L} 1 / |B| \sum\limits_{b \in B} \mathcal{S}_l (b) $ where $\mathcal{S}_l(b)$ interpolates the $l$-th CNN layer features from the 27 voxels surrounding the Euclidean coordinate $b$; $\Vert$ represents feature concatenation from $L$ layers.
The interpolated node features are used as the GNN inputs.
We use four graph attention layers (GAT)~\cite{gat} to obtain the latent graph representation and one GAT layer to obtain the final output layer.
The features of the $i$-th node $h^{\prime}_i$ of the GAT layers are computed as:
\begin{equation}
    h^{\prime}_i = \Vert_{k=1}^K \sigma ( \sum_{j \in \mathcal{N}_i} \alpha_{ij}^k \mathbf{W}^k h_{j} )
\end{equation}
where $\mathcal{N}_i$ is the first order neighborhood of node $i$ in the graph; $W^k$ is input linear transformation’s weight matrix corresponding to the $k$-th attention head; $\parallel$ represents concatenation; $\sigma$ is the ReLU activation. The attention coefficients $\alpha_{i,j}^k$ are computed as:
\begin{equation}
\alpha_{ij} =
\frac{
\exp\left(\mathrm{LeakyReLU}\left(a^{\top}
[\mathbf{W}h_i \, \Vert \, \mathbf{W}h_j]
\right)\right)}
{\sum_{k \in \mathcal{N}_i }
\exp \left( \mathrm{LeakyReLU} \left( a^{\top}
[\mathbf{W}h_i \, \Vert \, \mathbf{W}h_k]
\right)\right)}
\end{equation}
where $a$ is the weight vector of a single layer attention; $.^\top$ represents transposition.
The output layer is computed by averaging the $K$ attention heads instead of concatenation as: 
\begin{equation}
    h^{\prime}_i = sigmoid ( \frac{1}{K} \sum_{k=1}^K \sum_{j \in \mathcal{N}_i} \alpha_{ij}^k \mathbf{W}^k h_{j} )
\end{equation}
The ground-truth regression target is defined as the fraction of the branch segment that could be matched to the ground-truth tracing as $G(i) = \frac{1}{|B_i|} \sum\limits_{b \in B_i} t_b$ where $t_b = 1$ for all the nodes that are within a certain distance to the ground-truth centerlines and otherwise $t_b = 0$.
The binary cross entropy loss is computed between the output graph and the ground-truth target graph.
For inference, the branch segments $B_i$ with confidence below a threshold are discarded from the final result as visualized in Fig.~\ref{fig:gnn_prun_vis} (c) to (f).

\section{Experiment and Result}
We evaluated our proposed framework with 418 in-house multi-phase contrasted CT images. 
The images were be acquired with either arterial phase or venous phase. In both annotation and experiments, we did not differ different phases.
Each included image is ensured to cover the entire liver.
The liver vessels were traced by the annotators on intensity pre-processed 3D image volumes using Vaa3D \cite{vaa3d} and then refined by the certified radiologists.
We used 379 images for training and the rest 39 for testing.
All the images were spatially normalized to the resolution of $1^3$ mm.

To generate the training ground-truth centerline heatmap, $\alpha$ and $d_M$ are set as $6$ and $5$ respectively.
We used an initial learning rate of $8 \times 10^{-4}$ to optimize the vessel centerline enhancement CNN.
We used the open-sourced Rivuletpy package to implement the initial tracing algorithm~\cite{rivulet2} \footnote{https://github.com/lsqshr/rivuletpy}.
The graph attention layers were implemented using the PyTorch Geometric \cite{pytorch_geometric}.
We used Adam with a weight decay of $5\times10^{-4}$ and a learning rate of $5 \times 10^{-6}$ to optimize the GNN.
For GNN inference, we used a confidence threshold of $0.5$ to discard the false-positive branches.

To compare the result tracings against the ground-truth, we used the node catching based metrics, namely precision, recall, and F1 score, as well as the node distance-based metrics, namely spatial distance (SD), significant spatial distance (SSD) and percentile of significant spatial distance (pSSD) as used in~\cite{app2}.
We consider a predicted node caught by a ground-truth branch if they are within $4mm$.
Please note that the result tracings and the groundtruth here refer to the final reconstructed graph rather than the targets for GNN training.

In Table~\ref{tab:mnd}, we show that larger ground-truth matching distances for formulating the GNN objective loss increase the sensitivity while lowering the precision.
In Table~\ref{tab:sl}, we show the GNN performance for different choices of lengths to break the initial tracing graph into sub-branches.
We did not notice accuracy improvement for branch lengths longer than 5mm. We eventually fixed the node matching distance to 3mm and the branch sampling length to 5mm for the following comparisons.

\begin{figure}[!htb]
\centering
\includegraphics[width=0.95\linewidth]{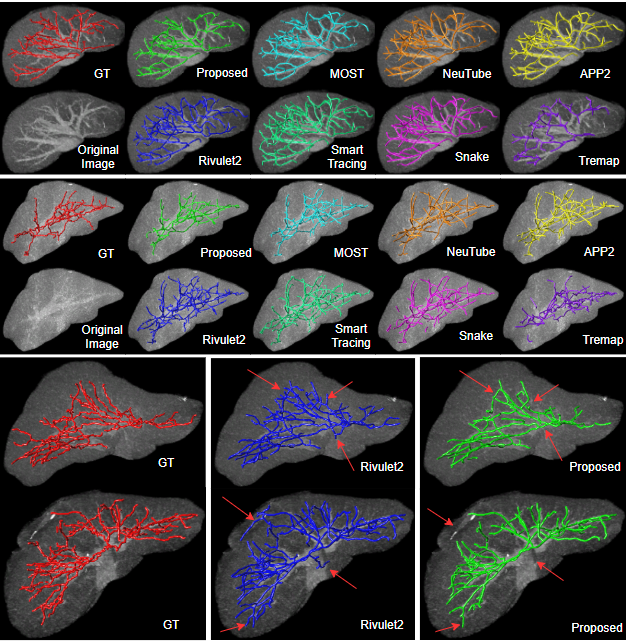}
\caption{Visual comparison with state-of-the-art methods. Red arrows indicate improvements from baseline Rivulet2 to the proposed method by removing false positive branches.}
\label{fig:exp_compare}
\end{figure}

\begin{table}[!htb]
\centering
\caption{Quantitative performance of different node matching distances (NMD).}
\resizebox{0.96\linewidth}{!}{%
\begin{tabular}{ccccccc}
\hline
NMD & Precision         & Recall           & F1               & SD                 & SSD              & pSSD \\ \hline
$15 mm$  & $0.785\pm0.094$   & $\textbf{0.893}\pm\textbf{0.065}$  & $0.830\pm0.049$  & $2.76 \pm 0.757$   & $9.62\pm2.53$  & $0.175\pm0.047$\\
$11 mm$  & $0.803\pm0.084$   & $0.892\pm0.066$  & $0.840\pm0.045$  & $2.65 \pm 0.731$   & $9.38\pm2.54$  & $0.166\pm0.043$\\
$7 mm$   & $0.866\pm0.058$   & $0.879\pm0.067$  & $0.8686\pm0.040$  & $\textbf{2.42} \pm \textbf{0.703}$   & $\textbf{9.35}\pm\textbf{2.77}$  & $0.143\pm0.037$\\
$3 mm$   & $\textbf{0.924} \pm \textbf{0.043}$  & $0.826\pm0.091$  & $\textbf{0.8688}\pm\textbf{0.056}$  & $2.49 \pm 0.735$   & $9.80\pm3.20$  & $\textbf{0.140}\pm\textbf{0.043}$\\ \hline
\end{tabular}}
\label{tab:mnd}
\end{table}

\begin{table}[!htb]
\centering
\caption{Quantitative performance of different branch sampling lengths.}
\resizebox{0.96\linewidth}{!}{%
\begin{tabular}{ccccccc}
\hline
SL & Precision         & Recall           & F1               & SD                 & SSD              & pSSD \\ \hline
$20 mm$ & $0.915\pm0.044$   & $0.818\pm0.089$  & $0.860\pm0.056$  & $2.54 \pm 0.590$   & $9.62\pm2.22$  & $0.146\pm0.054$\\
$15 mm$ & $0.918\pm0.049$   & $0.818\pm0.093$  & $0.861\pm0.059$  & $2.51 \pm 0.630$   & $\textbf{9.40}\pm\textbf{1.84}$  & $0.145\pm0.040$\\
$10 mm$ & $0.924\pm0.043$   & $0.826\pm0.091$  & $0.869\pm0.056$  & $2.49 \pm 0.735$   & $9.80\pm3.20$  & $0.140\pm0.043$\\
$5 mm$  & $\textbf{0.928}\pm\textbf{0.050}$   & $\textbf{0.835}\pm\textbf{0.085}$  & $\textbf{0.876}\pm\textbf{0.055}$  & $\textbf{2.46} \pm \textbf{0.668}$   & $9.72\pm3.17$  & $\textbf{0.136}\pm\textbf{0.043}$\\ \hline
\end{tabular}}
\label{tab:sl}
\end{table}

\begin{table}[!htb]
\centering
\caption{The quantitative comparison with state-of-the-art methods.}
\resizebox{0.97\linewidth}{!}{%
\begin{tabular}{ccccccc}
\hline
Method       & Precision              & Recall                   & F1                       & SD                    & SSD                    & pSSD   \\ \hline
MOST~\cite{most}          & $ 0.8625 \pm 0.0775$   & $0.6044  \pm 0.1330$     & $0.6984  \pm 0.0957$     & $2.97\pm1.196$        & $9.90 \pm 3.280$       & $0.167\pm0.0511$\\
NeuTube~\cite{neutube}       & $ 0.7622 \pm 0.1089$   & $0.2937  \pm 0.0703$     & $0.4158 \pm 0.0737$      & $3.86\pm1.116$        & $\textbf{9.28} \pm \textbf{2.517}$       & $0.279\pm0.0587$\\
APP2~\cite{app2}          & $ 0.8051 \pm 0.1531$   & $0.1890  \pm 0.1265$     & $0.2921 \pm 0.1629$      & $9.93\pm8.470$        & $16.44\pm 9.567$       & $0.348\pm0.1473$\\
Smart Tracing~\cite{smarttracing} & $ 0.7233 \pm 0.1487$   & $0.7780  \pm 0.2048$     & $0.7521 \pm 0.0771$      & $6.23\pm6.873$        & $14.87\pm 8.926$       & $0.255\pm0.1381$\\
Snake~\cite{snake}         & $ 0.7949 \pm 0.0870$   & $0.7719  \pm 0.1119$     & $0.7738 \pm 0.0616$      & $2.99\pm0.883$        & $10.34\pm 2.459$       & $0.174\pm0.0475$\\
TreMap~\cite{tremap}        & $ 0.7844 \pm 0.1610$   & $0.2287  \pm 0.1114$     & $0.3422 \pm 0.1313$      & $5.22\pm2.343$        & $9.40 \pm 2.974$       & $0.337\pm0.0920$\\
Rivulet2~\cite{rivulet2}      & $ 0.7634 \pm 0.0961$   & $\textbf{0.8823}  \pm \textbf{0.0678}$     & $0.8124 \pm 0.0514$      & $3.02\pm1.008$        & $10.27\pm 3.204$       & $0.184\pm0.0501$\\
Proposed      & $ \textbf{0.9280} \pm \textbf{0.0497}$   & $0.8354  \pm 0.0849$     & $\textbf{0.8762}  \pm \textbf{0.0549}$     & $\textbf{2.46} \pm \textbf{0.668}$        & $9.72 \pm 3.172$       & $\textbf{0.136}\pm\textbf{0.0433}$\\ \hline
\end{tabular}}
\label{tab:overallv2}
\end{table}

The same CNN produced vessel centerline heatmaps are used as the inputs for all the compared methods below.
The qualitative comparison with the state-of-the-art tracing methods is shown in Fig.~\ref{fig:exp_compare}.
It is shown that (1) the baseline Rivulet2 algorithm can produce generally better-initialized graphs than the other methods and (2) many false-positive branches traced by the baseline Rivulet2 were successfully pruned by the proposed method. 
We show the quantitative comparisons in Table~\ref{tab:overallv2}. 
NeuTube, APP2, and TreMap all under-reconstructed the graphs with a lower recall due to the gaps in the vessel heatmaps.
The adapted Rivulet2 algorithm produced the highest recall and the best F1 score comparing to the other tracing algorithms. We thus used Rivulet2 as the method for initializing the pre-pruning graphs.
The proposed method achieved the highest f1-score $0.8761$ with an increase of $6.38\%$ comparing to the baseline Rivulet2.
The increase in F1 score can be attributed to the $16.46\%$ precision increase at a cost of a slight drop of recall of $4.69\%$. It is notable that the proposed method still has higher recall than the other tracing algorithms except the baseline even after pruning.
The proposed method also achieved better SD, SSD, and pSSD than the baseline. The proposed approach is generic and adaptable to other 3D curvilinear graph tracing problems such as lung airway tracing, coronary vessel tracing and single neuron tracing. 

\section{Conclusion}
The morphology and topology of liver vascular structure is important for building a biological liver model for visualization and anatomy education. Liver vessel tree extracting is still challenging due to reasons such as morphological variability of the liver. In this work, we proposed a framework to reconstruct the 3D morphology models of the liver vessel tree from multi-phase CT images. The proposed framework uses a GNN to prune the false-positive branches generated by an image processing based tracing algorithm.
We evaluated the proposed method on a large-scale in-house 3D abdomen multi-phase contrasted CT image dataset, on which the proposed method outperformed state-of-the-art curvilinear reconstruction methods as well as improving the baseline method without GNN pruning by a large margin in terms of the tracing F1 score.

\noindent\textbf{Disclaimer}: The concepts and information presented in this paper are based on research results that are not commercially available.

\bibliographystyle{splncs04}
\bibliography{miccai20.bib}
\end{document}